      [1999/12/01 v1.4c Il Nuovo Cimento]
\documentclass{cimento}
\usepackage{graphicx}  
\usepackage{epsfig}
\title{Astrometry and Relativity}
\author{Costantino Sigismondi\from{ins:x}}
\instlist{\inst{ins:x} ICRA, International Center for Relativistic 
Astrophysics, and University of Rome La Sapienza, Piazzale Aldo Moro 5 00185 Roma, Italy.}
\PACSes{\PACSit{95.30.S}{Relativistic Effects}
\PACSit{95.10.Jk}{Astrometry}
\PACSit{98.62.Tc}{Astrometry}}
\begin{document}

\maketitle

\begin{abstract}
General relativistic effects in astrophyiscal systems have been detected thanks to accurate astrometric measurements. 
We outline some keystones of astrometry such as stellar aberration (argument development during the years 1727-1872); 
Mercury's perihelion precession (1845-1916); solar disk oblateness (1966-2001); 
relativistic light deflection (1916-1919); lunar geodetic precession (1916-1988); 
Lense-Thirring and Pugh-Schiff precessions (1917-1959), 
finally presenting the issue of the quest for a guide star for GP-B satellite (1974-2004) 
as application of all previous topics.
\end{abstract}

\section{Stellar aberration}

It is a relativistic effect, discovered in 1727 and well explained in a Galilean context. 
James Bradley discovered it looking for stellar parallaxes. 
He found that all the stars during a year describe an ellipse with semi-major axis of 20 arcseconds. 
Galilean explanation is straightforward looking at figure 1: when the Earth is in conjunction with the Sun with respect to a point of view external to the Earth's orbit, the Star appears to be in quadrature with the direction in space drawn dy this external point and the Earth. 
It occurs because of the Galilean composition of velocities. 

\begin{figure}
\centerline{\epsfxsize=4.1in\epsfbox{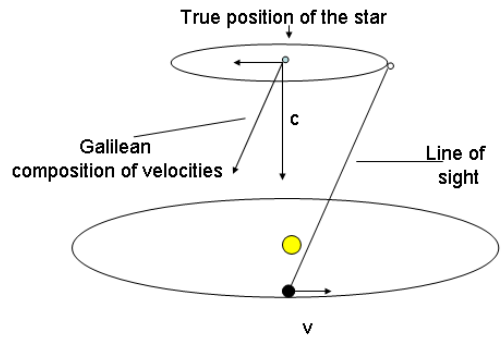}}      
\caption{Galilean composition of velocities in stellar aberration. The star is near the pole of the Ecliptic in order to magnify the effect.}
\end{figure}

Regardless of the moduli of the vectorial sums, the directions of the vectors are in agreement with the provenance of starlight. 
James Bradley in 1727 was looking for stars' parallaxes with respect to background stars (it was actually a Galileo's idea, strongly supported by Kepler, in order to prove the Copernican theory of an orbiting Earth). 
He expected to see, when the Earth was in quadrature with respect to a given point of the orbit like $\gamma$  point (i.e. a given direction of space), 
the star in quadrature with respect to its position with Earth at $\gamma$ point on the opposite side: 
a phenomenon which remained unobserved until 1838 (W. Bessel on $61~Cyg$ with the Heliometer of Fraunhofer, see figures 3 and 4) because of the smallness of the effect ($< 1$ arcsecond).

\begin{figure}
\centerline{\epsfxsize=4.1in\epsfbox{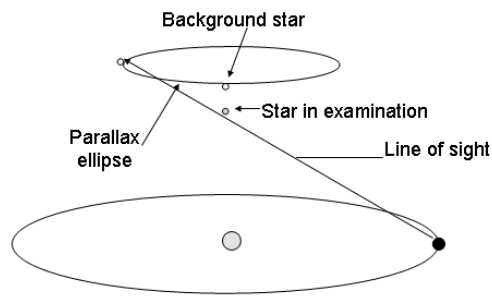}}
\caption{Parallax ellipse. The phase of parallax displacement is to be compared with
figure 1 of Galilean composition of velocities, their phases are separated of $\pi/2$.}
\end{figure}
 
\begin{figure}
\centerline{\epsfxsize=4.1in\epsfbox{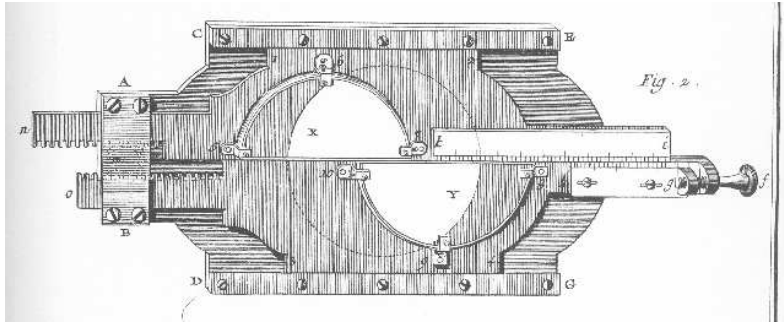}}
\caption{Heliometer. Scheme of the objective lens. Lens is splitted in two halves,  which can be moved with a micrometer. They produce two equal images splitted by a quantity depending on the displacement of the two half lenses. The comparison of the position of the star under examination with background stars
is made shifting it near them with the micrometer.}
\end{figure}

\begin{figure}
\centerline{\epsfxsize=4.1in\epsfbox{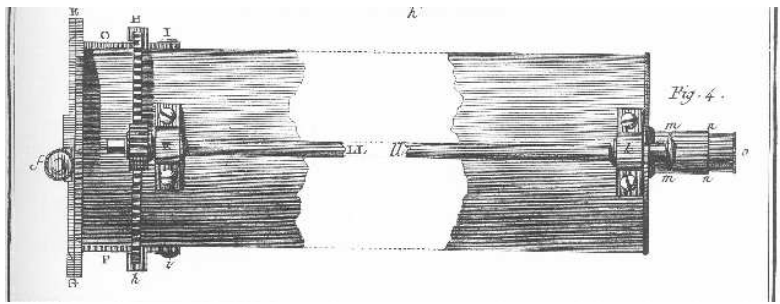}}
\caption{Apparatus in order to keep stable the focal lenght of the telescope with the Heliometer.}
\end{figure}

Aberration ellipse depends on the ecliptical coordinates of the star: the semi-major axis is $a=v/c=20$ arcsecs, where $v$ is the orbital velocity of the Earth. 
The semi-minor axis is approximately $b=v/c \cdot sin(\beta)$, with $\beta$ ecliptical latitude of the star.
The maximum displacement due to aberration occurs 3 months before the expected parallax effect with respect to background stars.

In the Galilean treatment 

$\tan(\theta-\theta')=v\cdot sin(\theta)/(c+ v \cdot sin(\theta))$

or expanding in Taylor series

$\theta ' = \theta - v/c \cdot (sin(\theta)- 1/2 v/c \cdot sin(2\theta)+$\ldots$)$

While in relativistic treatment 

$tan(\theta ')=tan(\theta)/(1+v \cdot sec(\theta)/c) \cdot (1-v^2/c^2)^{1/2}$ 

Phase and group velocity are the same in all inertial frames (for Galilean transformations there is no aberration 
in phase velocity because the angle of wavefront is an invariant).
There came out a question: since stellar light passes through Earth's atmosphere with refraction index $n$: which velocity is to be used, $c$ or $c/n$? 
In 1872 Airy  measured aberration with a telescope filled by water and published the results in the {\it Proceedings
of the Royal Society of  London}: the conclusion was that $c$ is to be used 
for aberration and the atmosphere is solidal with Earth \cite{ref:bar,ref:hos}.

\section{Mercury's perihelion advancement}

First attempts were addressed to explain this advancement with Newtonian planetary perturbations. 
Let us consider as an exemple what happens after a Mercury's orbit which lasts $\sim 1/4$ year.

\begin{figure}
\centerline{\epsfxsize=4.1in\epsfbox{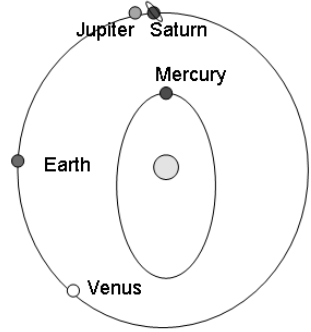}}
\caption{Planetary angular displacements after one orbit of Mercury. In this figure all planets start from uppermost position, moving counterclockwise.}
\end{figure}

The net perturbation of Earth after one orbit of Mercury is 
 $F_{\oplus} /F_{\odot} \sim (m_{\oplus}/m_{\odot})  \cdot  (r_{Mercury}/r_{\oplus})^2 \sim 1/200000$ 
corresponding to a perturbation of 1 arcsec per orbit ($\sim 200000$ arcsecs), i.e. 400 arcsec per century 
(400 orbits of Mercury per century). 
But in the case of Earth's Newtonian perturbations, 
after 4 Mercury's orbits, the net balance of Earth's perturbations almost cancels because of the stability of resonance. 

\begin{figure}
\centerline{\epsfxsize=4.1in\epsfbox{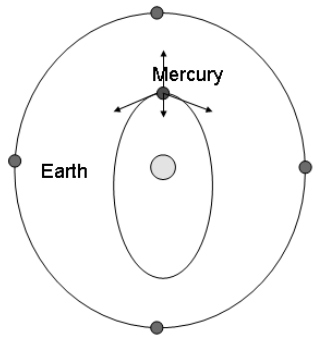}}
\caption{Stability of resonance between Mercury and Earth's orbits. Vectors representing gravitational forces acting upon Mercury, and Earth, are drawn at intervals of one orbit of Mercury.}
\end{figure}

Resonances with all planets are not all exact and a net perturbation arises. 
Its order of magnitude results the same of Earth's contribution after one orbit, i.e. $\sim 400$ arcsecs per century. 	
According to Le Verrier \cite{ref:flam} and Newcomb's \cite{ref:new,ref:moy} observations (1859-1882) all planetary perturbations yield 
an observed advancement of the perihelion of Mercury's orbit of $574.10\pm0.41$ arcsec per century. 
$42.56\pm0.5$ of them remain unexplained by Newtonian theory of gravitation. 
The reference frame for this advancement is also in motion due to the equinox (lunisolar) precession 
(discovered by Ipparchus $\sim$ 150 b.C.) of 50 arcsec per year i.e.~5000 arcsec/cy: it is a motion of the 
Earth's axis i.e. the celestial pole with respect to the ecliptic pole.
 
With this motion of the reference frame we include also Nnutation, due to the Moon's influence, another serendipitous discovery (during years 1727-1745) of J. Bradley 
on $\gamma$ Draconis whose declination oscillated of $\pm 18$ arcsecs over 18.6 years of observations.

\begin{table}
  \caption{Planetary perturbations for Mercury. \cite{ref:scia}}
  \label{tab:Plan}
  \begin{tabular}{rcl}
    \hline
      Perturbator    & Perturbation [arcsec/cy]  & errorbar [arcsec/cy]   \\
      Venus & 227.856 & 0.27    \\
      Earth & 90.038 & 0.08    \\
      Mars & 2.5536 & 0.00    \\
      Jupiter & 153.584 & 0.00    \\
      Saturn & 7.302 & 0.01    \\
      Uranus & 0.141 & 0.00    \\
      Neptune & 0.042 & 0.00    \\
      Solar Oblateness & 0.010 & 0.02    \\
      Total Newtonian & 574.069 & 0.30    \\
    \hline
  \end{tabular}
\end{table}

To explain the remaining $42.56 \pm 0.5$ arcsec/cy within Newtonian theory of gravitation have been considered: 

a) the perturbations of an intramercurial planet, Vulcan;

b) the effects of a small quadrupole moment of the Sun,  yielding a rosette-like orbit with advancing perihelion.

From an observative point of view: better observations of Mercury can be obtained when it is at the western or 
eastern elongation from the Sun, but in this case the motion of the planet is along the line of sight, and there are great errors 
on orbital elements estimation.

\begin{figure}
\centerline{\epsfxsize=4.1in\epsfbox{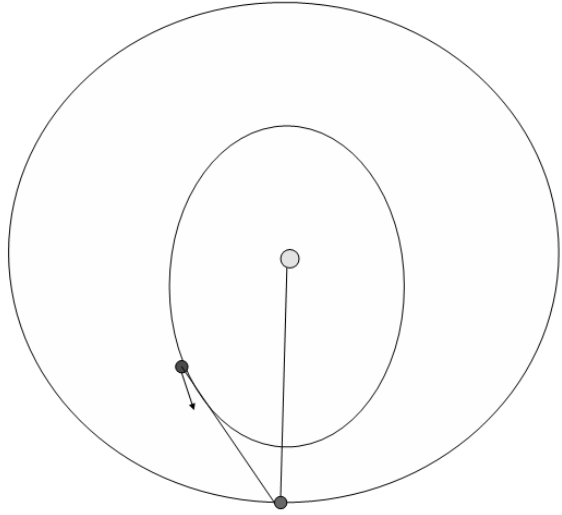}}
\caption{Measurement of orbital elements for Mercury: observative constraints. Mercury is better visible at its maximal elongations, but its velocity $v$ at that time is almost always along the line of sight. Transits across the solar disk provide better conditions for $v$ measurement.}
\end{figure}

Valuable observations are made during transits (last one on May, 7 2003) when the motion is perpendicular 
to the line of sight.
Observations from 1765 to now yield an anomalous precession of 
$43.1 \pm 0.1$ arcsec/cy explained by General Relativity as shown in table III \cite{ref:scia}. 

Einstein equations fully explain the anomalous precession of the perihelion of Mercury \cite{ref:ein} 
and of the other planets.

$\delta\theta=6\pi \cdot G M_{\odot} a/c^2b^2$ 

with a,b semiaxes of ellipse $b=a \cdot (1-e^2)^{1/2}$, 
e= eccentricity of the orbit.

\begin{table}
  \caption{General relativistic perturbations for Mercury. 
Third column shows the relative observability $e\cdot\delta\theta$: 
it is a parameter indicating that the more the orbit is
elliptic the more is detectable the perhielion advancement \cite{ref:scia}.}
  \label{tab:Plan2}
  \begin{tabular}{rcl}
    \hline
      Planet    & Perturbation [arcsec/cy]  & $e\cdot\delta\theta$ [arcsec/cy]    \\
      Mercury & 43.03 & 8.847    \\
      Venus & 8.63 & 0.059    \\
      Earth & 3.84 & 0.064    \\
      Mars & 1.35 & 0.126    \\
      Jupiter & 0.06 & 0.003    \\
    \hline
  \end{tabular}
\end{table}

Observations confirm Einstein predictions for the advancements of planetary perihelia.
Since   $\delta\theta \propto M_{\odot} /r$, with $r$ orbital distance,  this effect rapidly vanishes for planets far from the Sun.
Note that for orbits around Earth,  
 
$\delta\theta_{\oplus}/ \delta\theta_{\odot} = (M_{\oplus}/M_{\odot}) \cdot r_{\odot}/r_{\oplus}$$\sim ~ 1/33$ of the Mercury's value, for the 

closest orbit around Earth. Here $r_{\oplus}= 6578 $km i.e. $200$ km above Earth surface, while $r_{\odot}=50\cdot 10^{6}$ km is the orbital radius of Mercury.
For this reason the Moon shows a relativistic precession 1/2200 smaller than Mercury.

\begin{table}
  \caption{Relativistic perturbations: comparison with observations. \cite{ref:scia}}
  \label{tab:Plan3}
  \begin{tabular}{rcl}
    \hline
      Planet    & Perturbation [arcsec/cy]  & Observations [arcsec/cy]    \\
      Mercury & 43.03 & $43.1 \pm 0.1$     \\
      Venus & 8.63 & 8.65   \\
      Earth & 3.84 & 3.85 or $4.6\pm 2.7$\cite{ref:scia} \\
      Mars & 1.35 & 1.36    \\
      Moon & 0.02 & -   \\
    \hline
  \end{tabular}
\end{table}

\section{Solar oblateness}

There is a Netwonian precession in a quadrupole potential. Equation of quadrupole precession:
$\Omega_q=-3/2 \bar{\omega}_0 \cdot  (R/r)^2 \cdot cos(i)/(1-e^2)^2  \cdot J_2$,  where
$J_2=-Q_{33}/2MR^3 $ is an adimensional parameter for quadrupole moment, R the solar radius and $r$ is the orbital semiaxis, $\bar{\omega}_0$ the mean motion and $i$ the inclination of the orbit with respect to the equatorial plane \cite{ref:ciufo}.
If $ J_2=10^{-7} $ for the Sun (as from mass, rotation period and solar radius), the contribution to the precession experienced by Mercury should be $0.02$ arcsec/cy. See table IV for comparison with Earth's case.

\begin{table}
  \caption{Precession due to Solar oblateness ($J_2=10^{-7}$) and to Earth's one $J_{\oplus}\sim 1.083\cdot 10^{-3}$}
  \label{tab:Sol1}
  \begin{tabular}{rcl}
    \hline
      Planet    & Expected precession [arcsec/cy] \\
      Mercury & 0.02      \\
      Earth Oblateness & Around Earth satellite orbits are subjected to the quadrupole precession\\
        & \bf{{$\vec{d\Omega_q}/dt$}} $\propto -3/2 \bar{\omega}_0 [R_\oplus /a /(1-e^2)]^2 \cdot J_{2\oplus} \cdot cos(i) \cdot$ \bf{\^{n}$_{\oplus}$} \\
        & i, inclination of the orbit, $\bar{\omega}_0 $ satellite mean motion, $a$ radius of the orbit. \\
     \hline
  \end{tabular}
\end{table}

\subsection{Solar disk astrometry}

Several Earthbased experiments have been conducted to measure the solar disk ellipticity. Main
astrometric problems from the Earth are: 

1) Astronomical refraction which is responsible of a non circular shape of the Sun, especially near the horizon. It was discovered by Tycho Brahe studying the Supernova of 1571. 
For moderate zenithal distances z, the increasement of height $\delta\Theta$ above the horizon of a point of the 
solar disk is

$\delta\Theta \sim 60" \cdot tan(z)$. 

The value of $\delta\Theta$ approaches the limit of $\delta\Theta \approx 34$ arcminutes near horizon. The previous formula no longer applies and Garstand's fit for airmass versus z is to be used \cite{ref:gars}.
Since the sun is still visible when its true upper limb is $34$ arcminutes below the horizon the duration of the day is longer than that one calculated without astronomical refraction \cite{ref:duff}. 
At horizon the figure of the Sun appears elliptical with vertical semiaxis smaller than the horizontal one up to 6 arcminutes (apparent oblateness). 

\begin{figure}
\centerline{\epsfxsize=4.1in\epsfbox{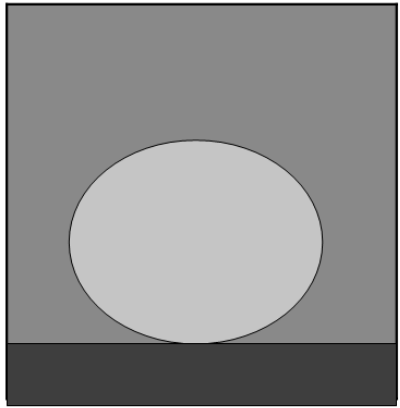}}
\caption{Solar disk apparent oblateness at sunset or sunrise.}
\end{figure}

2) Light aberration, which produces an effect along the solar equator, due to the rotation of the Sun around its axis.

3)Horizontal deformation, due to Earth's rotation. Thiss another aberration effect due to the Galilean composition of Earth's rotation velocity with respect to the Sun $v_{\odot}=v \cdot cos(\lambda)\cdot cos(a)$  and the speed of light $c$, with $v=0.46$ km/s, $\lambda$ the latitude, $a$ the azimuth (a=0 when the Sun transits on the local meridian).

The effect is slightly different from eastern to western solar limb because they are at different azimuth $a$, and it produces a small deformation on the horizontal direction.

\subsection{Solar Disk Sextant}

It is the most recent experiment on the measurement of the Sun \cite{ref:sof}.
A rotating telescope above the atmosphere takes the positions of 10 points of the solar disk.
Large photon statistics allow the precise location of those points. 
After data reduction for aberration and optical distortions the expected errorbar on the solar diameter is few milliarcseconds. 
the goal of this experiment is to detect secular variations of the solar diameter, beyond the $11$-year sunspots' cycle. 

\begin{figure}
\centerline{\epsfxsize=4.1in\epsfbox{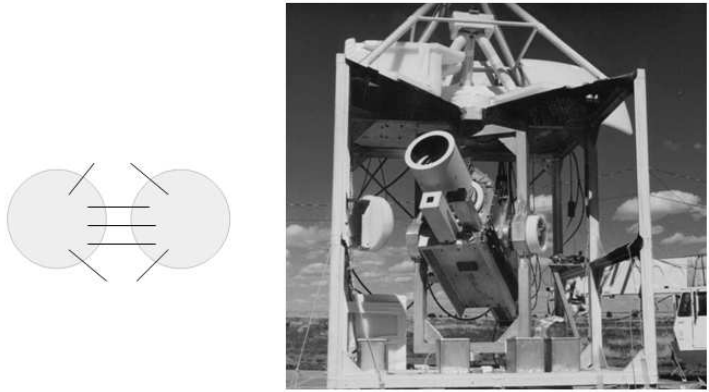}}
\caption{Solar Disk Sextant's focal plane CCD configuration. SDS is a Yale-NASA project to which the author has participated.}
\end{figure}

\section{Relativistic light deflection}

John Michell (1784) considering light as corpuscular, conceived the idea of a gravitating light and therefore of a black hole \cite{ref:eis}.

In general relativity $\Delta E = + 0".0047 \cdot 1/tan(E/2)$  where E is the elongation of the star from the center of the Sun ($\Delta \Theta=4GM_{\odot}/bc^2$). 
Remarkable is the effect done by Sun on the light coming from $Iades$ cluster in occasion of the total solar eclipse of 1919 \cite{ref:edd}.
The Earth yields a similar 
$\Delta \Theta_{\oplus}=(10^6/12 \cdot 10^5)\cdot \Delta \Theta_{\odot}$, for an electromagnetic signal coming from a satellite orbiting at 600 km of altitude.

\begin{figure}
\centerline{\epsfxsize=4.1in\epsfbox{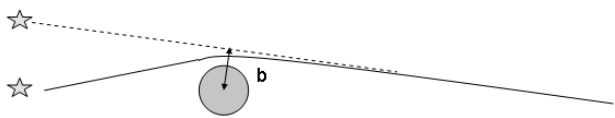}}
\caption{Relativistic light deflection by the solar mass.}
\end{figure}

\section{Geodetic (de Sitter) Precession}

Parallel transport (of a constant spin vector) in curved spacetime along a geodetic line 
(an orbiting body is actually in free fall, therefore along a geodetic line) generates a precession with respect to a fixed reference frame.

$\vec{\Omega}_{dS} = 3/2 \cdot GM_{\odot}/c^2r^3 _{\odot}\cdot$ ($\vec{r}\wedge\vec{v}$).

This precession has been predicted by Wilhelm de Sitter \cite{ref:des}.
For a spinning satellite at 600 km of altitude around Earth, after one orbit, such a precession is  (for a circular orbit after applying third Kepler law) 
$\Theta_{dS}=3/2·GM_{\oplus}/c^2r^2 \cdot (vGM_{\oplus}/r_{\oplus})^{1/2} \sim ~6.6$ arcsecs.

De Sitter precession depends on parallel transport in curved spacetime along geodetics,  while
Thomas precession occurs in flat spacetime (special relativity) with accelerated bodies (non geodetic motion).
Thomas precession in General Relativity occurs when additional non gravitational strengths deviate the body from geodetic motion \cite{ref:schiff}. 

\subsection{de Sitter precession of the Moon's orbit}

After one Earth's orbit such a precession is  (again calculated for a circular orbit) 
$\Theta_{dS}=3/2·GM_{\odot}/c^2r^2 \cdot (vGM_{\odot}/r_{\odot})^{1/2} \sim ~0.0192$ arcsecs.
This precession has been measured within $2\%$ of accuracy by Bertotti et al. (1987)\cite{ref:ber}.
Note that this precession is along the direction of the motion.
After one orbit the "spin vector" (in this case the orbital angular momentum of the Moon around Earth) precesses in the direction of the  orbital motion.

\begin{figure}
\centerline{\epsfxsize=4.1in\epsfbox{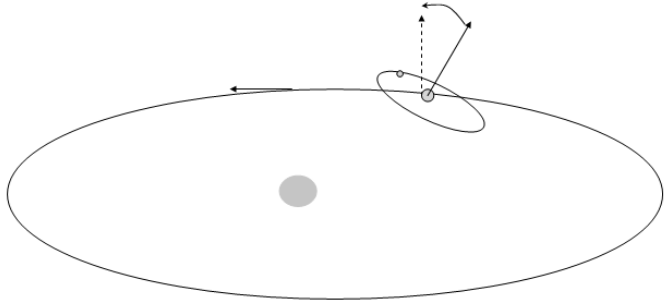}}
\caption{Scheme of de Sitter precession in the case of Moon's orbital spin around the Sun.}
\end{figure}

\subsection{Relativistic precessions as coupling between angular momenta}

Rewriting de Sitter precession formula as coupling between spin and orbital angular momenta

 $d\vec{S}/dt=3/2 \cdot GM/mc^2 r^3 \cdot~(\vec{L}\wedge \vec{S})$

where,  in the previous case, $\vec{S}$ is the orbital momentum of the Moon, or  a constant spinning vector,  and
$\vec{L}$ is the orbital momentum of the Earth-Moon system, or -in general- that one of the body carrying the constant spinning vector.

\section{Lense-Thirring precession}

A constant spin vector orbiting around a spinning body, since rotational energy modifies spacetime, is subjected also to a relativistic spin-orbit coupling, which drags the orbiting body out of the original orbital plane.
The angular velocity vector of Lense-Thirring precession is

$\vec{\Omega}_{LT}=GI_{\oplus}/2c^2/R^3_{\oplus} \cdot [3\vec{R}/R^2_{\oplus} (\vec{\omega} \cdot \vec{R})-\vec{\omega}]$,
here $R_{\oplus}$ is the Earth's radius, and $\vec{R}$ 
is the position (vector) of the orbiting gyroscope; I and $\vec{\omega}$ the moment of inertia and angular velocity vector of the Earth [see Lense-Thirring papers reproduced in \cite{ref:sigi}] . 

\subsection{Lense-Thirring vs perihelion precession}

In 1917-18 Hans Thirring asked the astronomer Josef Lense to help him in calculating the effects of gravitational field around a rotating mass.
For Mercury they found a precession $0.01~ arcsecs/cy$ in the direction opposite to the rotation of the Sun, or to the orbital motion. 
For a polar orbit the precession (of apsidal line) occurs in the same direction of the rotation of the central mass, and a consequence of that is the changement of the original orbital plane (see figure 12).
This effect is a consequence of general relativistic equation of motion and off-diagonal space-time components of the metric tensor which cannot be inferred by equivalence principle \cite{ref:schiff}. 
General Relativity predicts conditions under which the first Kepler Law, of planar and elliptic motion for two-body gravitational interaction, is no longer valid.

\subsection{Lense-Thirring torques}

For an eastward rotating central mass, a body in polar orbit is subjected to an average torque  eastward, and its orbital plane changes. $< \vec{\Omega_{LT}} > \propto < \vec{\omega} >$ (here $< \vec{\Omega} >$ and $< \vec{\omega} >$ are both average vectors).
For an equatorial orbit the torque is on the plane and westward and $< \vec{\Omega_{LT}} > \propto - < \vec{\omega} >$.

$d\vec{L}/dt=(\vec{\Omega} \wedge \vec{L})$ rules the evolution of orbital angular momentum $\vec{L}$, precessing around the vector $\vec{\Omega}_{LT}$, Newtonian plane orbits (first Kepler's law) are changed into precessing \textit{spherical orbits}.  

\begin{figure}
\centerline{\epsfxsize=4.1in\epsfbox{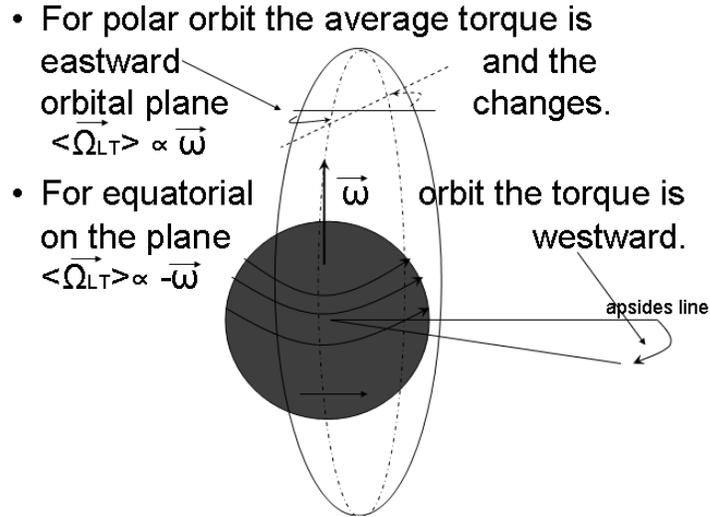}}
\caption{Lens-Thirring precession: scheme of the direction of the relativistic torque.}
\end{figure}

If we think to the rotating central body as "dragging" the metric with it, and we test the metrtic with an orbiting and spinning gyroscope, near the poles there is a tendedncy for the metric to rotate with the central body. Therefore a spin which is orbitating around that body precess in the direction of the rotating body. While near the equator the gravitational field and also the "dragging" of the metric falls off with increasing radial distance. If, then, we imagine the gyroscope, oriented so that its axis is perpendicular to that of the central rotating body, the side of the gyroscope nearest that body is dragged with the body more than the side away from it, so that the spin precesses in the opposite direction to the rotation of the body.

\subsection{Lense-Thirring orbits in Kerr field}

The solution of Lense and Thirring are perturbative solutions, valid at first order in case of large distances and low $\vec{\omega}$ of central rotating body.
More complicate orbits arise from positions near the horizon of a black hole or a neutron star, with whatever starting orbit.
There are spherical belts of allowance within which the orbits occur. Those belts are drawn by precessing lines of apsides. (Wilkins, 1972)

\begin{figure}
\centerline{\epsfxsize=4.1in\epsfbox{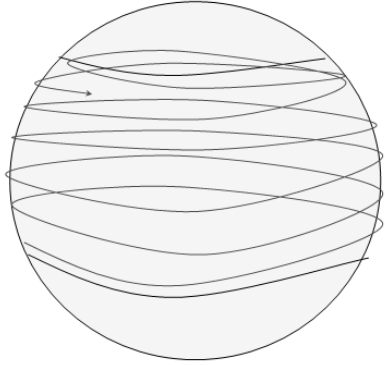}}
\caption{Sketch of a spherical orbit around a Kerr field. The body arises to a maximum latitude and turn back to a minimum one and so on.}
\end{figure}                     

\subsection{Higher order torques}
In the Lense-Thirring effect the spin of the central mass drags the orbital angular momentum $\vec{L}$.
If we consider the spin $\vec{s}$ of the orbiting gyroscope, it is also subjected to a smaller torque $\propto (s/L)\cdot \Omega_{LT}$. Also de Sitter term appears $\propto (s/L)\cdot \Omega_{dS}$

\section{Pugh-Schiff precession}
Instead of considering the precession of the lines of apsides of the elliptical orbit, the precession of the spinning axis of a torque-free gyroscope orbiting around Earth is studied for evidencing Lense-Thirring torques \cite{ref:schiff}. [Pugh's paper of 1959 has been reprinted in \cite{ref:sigi}]. Such a study has started the project for the Gravity Probe-B satellite which is scheduled for launch in 2004 \cite{ref:GPB}.

\begin{table}
  \caption{Lense-Thirring timing corrections for planets' satellites ephemerides [see Lense-Thirring papers in \cite{ref:sigi}]}
  \label{tab:Plan5}
  \begin{tabular}{rcl}
    \hline
      Satellite   & $\Delta T$ [s] after 100 years   \\
      Moon & 13.9    \\
      Phobos (Mars) & 0.5 \\
      Io (Jupiter) & 29.5 \\
      Amalthea (Jupiter) & 65.4 \\
      Mimas (Saturn) & 19.2    \\
      Ariel (Uranus) & 3.7  \\
    \hline
  \end{tabular}
\end{table}

\section{The guide star for GP-B}

Originally \cite{ref:eve} Rigel, a 0.7 magnitude star laying near the celestial equator, was choosen as referencee star for GP-B satellite. Afterwards IM Peg ($HR~8703$) of magnitude $M=5.9$ has been selected\cite{ref:sha}. 
It is a radio active star, close to a radio quasar. This has been done in order to measure very accurately its proper motion with VLBI, within $0.09$ milliarcseconds of accuracy.

A beam splitter produces two images of the reference star for each readout axis subjected to photon counting statistical errors.  Orbital light aberration ($\pm 5$ arcsecs and $90$ minutes of period); annual aberration ($\pm 20.116$ arcsecs) and light deflections for Sun and planets produce large, very accurately known, periodic displacements.
Those displacements appear in the readout of each gyroscope of GP-B, allowing a continuous calibration of the gyro scale of few parts in $10^7$, and a  precision of $0.1$ milliarcsecond over 3 years of experiment [see Everitt's paper in \cite{ref:sigi}].

\end{document}